# DAMPED LYMAN-ALPHA AND LYMAN LIMIT ABSORBERS IN THE COLD DARK MATTER MODEL

Neal Katz[1], David H. Weinberg[2,5], Lars Hernquist[3,4] Jordi Miralda-Escudé[5]

E-mail: nsk@astro.washington.edu, dhw@payne.mps.ohio-state.edu, lars@helios.ucsc.edu, jordi@sns.ias.edu

## ABSTRACT

We study the formation of damped Ly$\alpha$ and Lyman limit absorbers in a hierarchical clustering scenario using a gas dynamical simulation of an $\Omega = 1$ cold dark matter universe. In the simulation, these high column density systems are associated with forming galaxies. Damped Ly$\alpha$ absorption, $N_{\rm HI} \gtrsim 10^{20.2}$ cm$^{-2}$, arises along lines of sight that pass near the centers of relatively massive, dense protogalaxies. Lyman limit absorption, $10^{17}$ cm$^{-2} \lesssim N_{\rm HI} \lesssim 10^{20.2}$ cm$^{-2}$, develops on lines of sight that pass through the outer parts of such objects or near the centers of smaller protogalaxies. The number of Lyman limit systems is less than observed, while the number of damped Ly$\alpha$ systems is quite close to the observed abundance. Damped absorbers are typically $\sim 10$ kpc in radius, but the population has a large total cross section because the systems are much more numerous than present day $L_*$ galaxies. Our results demonstrate that high column density systems like those observed arise naturally in a hierarchical theory of galaxy formation and that it is now possible to study these absorbers directly from numerical simulations.

*Subject headings:* Methods: numerical, Hydrodynamics, Galaxies:formation, large-scale structure of Universe, quasars: absorption lines

[1]University of Washington, Department of Astronomy, Seattle, WA 98195

[2]Ohio State University, Department of Astronomy, Columbus, OH 43210

[3]University of California, Lick Observatory, Santa Cruz, CA 95064

[4]Sloan Fellow, Presidential Faculty Fellow

[5]Institute for Advanced Study, Princeton, NJ 08540

## 1. Introduction

Damped Ly$\alpha$ and Lyman limit absorption systems are among the most useful probes of highly nonlinear structure at moderate and high redshifts. They are much more common than quasars or luminous radio galaxies, and several lines of evidence suggest that they trace the population of "normal" galaxies in the young universe. The mass of atomic hydrogen in damped systems at $z \sim 3$ is similar to the mass of stars in spiral disks today (Wolfe 1988), and observations show that at least two damped systems have linear extents $\gtrsim 10h^{-1}$ kpc (Briggs et al. 1989; Wolfe et al. 1993). Deep imaging and spectroscopic studies indicate that Lyman limit systems lie on lines of sight that pass near bright galaxies (Yanny 1990; Lanzetta & Bowen 1990; Bergeron & Boissé 1991; Steidel, Dickinson & Persson 1994). Existing surveys have detected many Lyman limit absorbers (e.g., Sargent et al. 1989; Lanzetta 1991; Storrie-Lombardi et al. 1994; Stengler-Larrea et al. 1995) and damped Ly$\alpha$ systems (e.g., Wolfe et al. 1986; Lanzetta et al. 1991), and the Keck telescope will push absorption surveys to even greater redshifts (Wolfe et al. 1995).

Previous comparisons between observed properties of high column density absorbers and theories of structure formation have been based on simplified approximations. Mo & Miralda-Escudé (1994) and Kauffman & Charlot (1994) use the Press-Schechter (1974) formalism to predict the abundance of dark matter halos and assume that all gas within these halos cools and becomes neutral. Ma & Bertschinger (1994) and Klypin et al. (1995) use a similar approach, but they normalize their Press-Schechter fits with dissipationless N-body simulations. These results can be compared to the amount of neutral gas inferred from the observations, but this analytic approach does not yield other properties of the absorption systems, and it can easily underestimate the constraining power of the observational data because it is based on the optimistic assumption of perfectly efficient gas cooling within the halo virial radius.

In this paper we predict the absorption statistics of Lyman limit and damped Ly$\alpha$ systems directly from a gas dynamical simulation, thereby eliminating many of the uncertainties required by previous methods. Our results for the Ly$\alpha$ forest — lines with column densities below $10^{17}$ cm$^{-2}$ — are presented in a companion paper (Hernquist et al. 1995; hereafter HKWM).

## 2. The Simulation

We explore a standard cold dark matter (CDM) universe with $\Omega = 1$, $H_0 = 50$ km/sec/Mpc, $\Omega_b = 0.05$, and the perturbation spectrum normalized to $\sigma_{16} = 0.7$. We simulate a comoving periodic volume that is $22.22/(1 + z)$ Mpc on a side and include both radiative and Compton cooling for a gas of primordial abundance. Our initial conditions are identical to those of Katz, Hernquist & Weinberg (1992; hereafter KHW) and Hernquist, Katz & Weinberg (1995), but here the mass resolution is eight times higher: we use a total of 524,288 particles, half to represent the dark matter and half to represent the gas. The gas particle mass is $1.5 \times 10^8 M_\odot$ and the dark matter particle mass is $2.8 \times 10^9 M_\odot$. Another important difference from KHW is that we include a photoionizing background, with uniform intensity $J(\nu) = J_0 \left(\nu_0/\nu\right) F(z)$, where $\nu_0$ is the Lyman-limit frequency, $J_0 = 10^{-22} \, \mathrm{erg \, s^{-1} \, cm^{-2} \, sr^{-1} \, Hz^{-1}}$, and

$$F(z) = \begin{cases} 0, & \text{if } z > 6; \\ 4/(1+z), & \text{if } 3 \leq z \leq 6; \\ 1, & \text{if } 2 < z < 3 \, . \end{cases}$$

We compute abundances assuming ionization equilibrium and an optically thin gas, and we use these abundances when computing rates of radiative cooling and photoionization heating.

We originally performed this simulation to examine the impact of photoionization on the formation of low mass galaxies; the results of this study, and the details of the calculation, are described elsewhere (Weinberg, Hernquist & Katz 1995). Briefly, the simulation was performed with TreeSPH (Hernquist & Katz 1989), a Lagrangian code that combines smoothed particle hydrodynamics (SPH) with a hierarchical tree algorithm for computing gravitational forces. The cosmological version of TreeSPH and the methods used to compute radiative cooling in the presence of an ionizing background are described by Katz, Weinberg & Hernquist (1995, hereafter KWH).

When computing gravitational forces, we use a critical opening angle, $\theta = 0.7$ (*e.g.* Barnes & Hut 1986; Hernquist 1987). The equations of motion are integrated so that all the dark matter particles are moved on the system time step of $3.2 \times 10^6$ years. TreeSPH, allows SPH particles to have time steps smaller than those of the dark matter particles by powers of two, up to a factor of 16 smaller than the system time step. The gravitational resolution is 20 comoving kpc (13 kpc equivalent Plummer softening), and the gas resolution varies from 5 comoving kpc in the highest density regions to 200 comoving kpc in the lowest density regions, reflecting the variable Lagrangian nature of TreeSPH. Compared to KHW, the higher mass resolution makes the simulation much more computationally expensive, so we have evolved it only to $z = 2$, taking 300 CPU hours on a Cray C–90.

When its power spectrum is normalized to $\sigma_{16} = 0.7$, the $\Omega = 1$ CDM model reproduces the observed mass function of rich galaxy clusters reasonably well (White, Efstathiou & Frenk 1993). However, this normalization is nearly a factor of two below that implied by the COBE microwave background fluctuations. This disagreement between the cluster-scale and COBE normalizations for "standard" CDM is, in our opinion, the most serious observational difficulty for this model. This discrepancy can be resolved in a number of ways, each being somewhat contrived but retaining a critical density universe dominated by cold dark matter. Among these variations are "tilting" or "breaking" the inflationary fluctuation spectrum, lowering the Hubble constant to $H_0 \sim 30 \, \mathrm{km \, s^{-1} \, Mpc^{-1}}$, including gravity wave corrections expected in typical inflationary models, or introducing a decaying particle that boosts the radiation content of the universe, delaying the epoch of matter domination. Each of these modifications would lower the amplitude of cluster-scale fluctuations relative to COBE fluctuations, and each would yield a somewhat different shape for the power spectrum on the scale of our simulation. Instead of selecting arbitrarily between these possibilities, we opt to consider the standard CDM model with a sub-COBE normalization, so that we can compare our simulation to the wealth of earlier N-body and hydrodynamic studies of this scenario. We suspect that our results for the hydrogen absorption lines would translate without major modification to other $\Omega = 1$, CDM-dominated models that have a similar $\sigma_{16}$ normalization, but the sensitivity of the hydrogen absorption lines to the shape of the power spectrum between cluster and sub-galactic scales is clearly an important issue for future investigations.

To calculate the distribution in HI column densities of the absorption systems, we project the neutral hydrogen mass of each gas particle onto a 2-dimensional uniform grid using a cell spacing that is comparable to the highest resolution achieved anywhere in the simulation, about $5.4/(1 + z)$ kpc. In this paper we consider absorption features with column densities in excess of $10^{17}$ cm$^{-2}$. The probability that more than one of them lies along a line of sight through the simulation is very small, since at $z = 2$, 3, and 4 the redshift interval across the simulation box, $\Delta z$, is 0.019, 0.030, and 0.041, respectively, and, on average, there are observed to be only a few Lyman limit systems per unit redshift. To count systems with much smaller column densities one must construct artificial spectra, as in HKWM.

Throughout the course of a simulation, we assume that the gas is optically thin to the ionizing radiation. This is adequate to compute the radiative cooling because gas that is dense enough to self-shield cools rapidly even when optically thin abundances are used. However, the neutral fractions are needed for the absorption line analysis, and self-shielding should be important in increasing the column densities of systems with $N_{HI} > 10^{17}$cm$^{-2}$. In our analysis, we correct for self-shielding on a pixel-by-pixel basis, treating each absorber as a plane-parallel slab with a perpendicularly incident radiation field. While this is an idealized approximation, our conclusions are not sensitive to its details, which are described below.

For each pixel we first calculate the neutral hydrogen column, neutral mass-weighted temperature, and neutral mass-weighted neutral fraction in the optically thin approximation. Using the temperature and neutral fraction, we solve for the average 3-dimensional hydrogen density using a bisection algorithm, then determine the total hydrogen column using the neutral mass-weighted neutral fraction and the neutral hydrogen column. To compute the self-shielding correction we assume that the total hydrogen is distributed uniformly in a plane-parallel slab, with half of the background radiation field incident perpendicular to each side of the slab. These assumptions do not themselves represent a realistic absorption geometry, but they are intended to be intermediate between a true slab and a sphere in an isotropic radiation background.

We divide the slab into 100 equal computational bins. Starting at the outermost bin and sweeping through to the middle of the slab, we calculate new equilibrium abundances with the background field attenuated by HI, HeI, and HeII absorption in the other bins. However, since the abundances have now changed, the heating and cooling rates in the bin have also changed and hence there should be a new equilibrium temperature. We calculate the new equilibrium temperature including any adiabatic heating or cooling terms that are present. The new temperature implies new equilibrium abundances, so we repeat the calculation, iterating within each bin until its abundances converge. We then repeat the computation for the full slab (since the new abundances imply new attenuation factors) and iterate until every bin in the slab has converged.

At column densities of $10^{16}$ to $10^{17}$ cm$^{-2}$, the correction is small and varies between 1% and 10%. At higher column densities, the correction can be as large as a factor $\sim 100$. The self-shielding correction must be applied to over 200,000 pixels at each redshift, so we use the above procedure to create a 3-dimensional (column density, neutral fraction, temperature) lookup table and compute individual pixel corrections by interpolation.

## 3. Results

Figure 1 shows the projected neutral hydrogen map at $z = 2$. All lines of sight with column densities greater than $10^{17}$ cm$^{-2}$ appear white in Figure 1. These regions are well separated from one another. The interconnected, filamentary structures give rise to absorption at lower column densities. Even lines of sight through regions that appear black in Figure 1 can give rise to HI absorption at Ly$\alpha$ forest column densities (see HKWM).

We use the algorithm of Stadel et al. (1995; see also KHW, KWH) to identify gravitationally bound clumps of dense ($\rho/\bar{\rho} > 1000$), cold ($T < 30,000$K) gas. These sites are likely to harbor rapid star formation and can therefore be identified with forming galaxies. They range from massive, high overdensity systems ($\rho/\bar{\rho} \gtrsim 10^5$) containing hundreds of particles to lower overdensity clumps containing as few as 8 cold gas particles. The former are usually "mature" objects that began to form at $z \sim 3 - 6$, while the latter are young systems that have just begun to condense and cool. Figure 2 plots $D_{\rm proj}$, the projected separation between an absorber and the center of the nearest galaxy in the simulation, against neutral column density, at $z = 2$. We sample a roughly equal number of lines of sight in each decade of $N_{\rm HI}$. When $D_{\rm proj}$ exceeds 100 kpc, it is plotted at 110 kpc.

Figure 2 shows a clear association between high column density absorbers and forming galaxies. Damped Ly$\alpha$ absorption, with $N_{\rm HI} \geq 10^{20.2}$ cm$^{-2}$, occurs along lines of sight that pass through the denser, more massive protogalaxies. The column density correlates (inversely) with projected separation, and the maximum $D_{\rm proj}$ for damped absorption is about 20 kpc. Closer inspection of the neutral gas in these systems often shows evidence of flattening and rotational support, but our limited gravitational resolution of $20/(1+z)$ kpc prevents us from drawing strong conclusions about the morphology of these objects.

Lyman limit absorption ($10^{17}$ cm$^{-2} \leq N_{\rm HI} \leq 10^{20.2}$ cm$^{-2}$) occurs on lines of sight that pass either through the outer parts ($D_{\rm proj} \sim 20 - 100$ kpc) of the more massive protogalaxies or near the centers of younger, lower density systems. Some of these absorbers have $D_{\rm proj} > 100$ kpc, but these are mostly cases where the corresponding clump of cold gas has too few particles to be identified as a galaxy by our algorithm. If the simulation were repeated with better mass resolution, these lines of sight would probably have associated, low mass protogalaxies, and they might exhibit higher column density absorption because of denser gas and more efficient cooling.

Fig. 1.— Projected neutral hydrogen in the simulation at $z = 2$. The simulation box is 22.22 comoving Mpc across, which corresponds to 7.4 physical Mpc at this redshift. For the CDM cosmology used, this translates into 15.1 arc minutes. A self-shielding correction was applied to this HI map (see text for details). In the color scheme, white corresponds to $N_{\rm HI} \gtrsim 10^{16.5}$ cm$^{-2}$, yellow to $10^{15.5}$ cm$^{-2} \lesssim N_{\rm HI} \lesssim 10^{16.5}$ cm$^{-2}$, red to $10^{14.5}$ cm$^{-2} \lesssim N_{\rm HI} \lesssim 10^{15.5}$ cm$^{-2}$, and black to $N_{\rm HI} \lesssim 10^{14.5}$ cm$^{-2}$.

Figure 3 shows the HI column density distribution $f(N,z)$, the number of absorbers per unit interval of column density per unit "absorption distance" $\Delta X \equiv (1+z)^{1/2} \Delta z$ (we use the $\Omega = 1$ definition appropriate to our simulations instead of the $\Omega = 0$ definition used in many observational papers). Histograms are computed from the projected neutral hydrogen maps at $z = 2$, 3, and 4 by counting the fraction of pixels in each column density bin. The observational data for damped systems were kindly provided to us by K. Lanzetta (1995, private communication); earlier versions of these data were shown in slightly different form in Lanzetta (1991) and Wolfe et al. (1995). It is difficult to determine column densities for absorbers that are opaque in the Lyman continuum but not strong enough to produce damping wings. However, it is more straightforward to determine the cumulative number of systems with $N_{\rm HI} > 10^{17.2}$, i.e. with an optical depth of unity at the Lyman limit. The diagonal boxes in Figure 3 represent the results of Storrie-Lombardi et al. (1994), which we have converted from cumulative numbers to constraints on $f(N,z)$ by assuming that the form of $f(N)$ is the $N^{-1.32}$ power law found by Petitjean et al. (1993) in this regime of column density. The total redshift path length spanned by the $4096^2$ lines of sight through the simulation at the three different redshifts amounts to $\sim 1.5 \times 10^6$, over 1000 times the total path length covered by all existing observations, though our lines of sight are not all independent. The simulated frequency distribution shows minimal evolution between $z = 2$ and $z = 3$ and weak evolution between $z = 3$ and $z = 4$.

## 4. Discussion

For damped Ly$\alpha$ absorbers, the agreement between the simulated and observed $f(N,z)$ is fairly good. The simulation numbers are a bit low, but it is not clear that this difference represents a significant failure of the CDM model. While our simulation volume is randomly chosen, it may not be large enough to be statistically representative — a larger box would include larger scale modes that would lead to the formation of "proto-Coma" clusters and "proto-Böotes" voids in which the collapse of galaxy scale objects would be, respectively, accelerated or retarded. Higher numerical resolution would probably increase the amount of cooled gas and hence $f(N,z)$. However, star formation, neglected in this calculation, would reduce $f(N,z)$ by converting neutral gas into stars, perhaps even reversing the sign of the evolutionary trend in Figure 3 (see, e.g., Kauffmann & Charlot 1994, Wolfe et al. 1995). We will examine these effects in future studies.

Wolfe et al. (1995) estimate $\Omega_g$, the density parameter of mass in damped Ly$\alpha$ systems, by integrating $f(N,z)$ between $N_{\rm HI} = 10^{20.2}$ cm$^{-2}$ and $N_{\rm HI} = 10^{21.7}$ cm$^{-2}$, the highest column density at which they have observed systems. They find $\Omega_g = 0.002 \pm 0.001$ at $z = 2$, and $\Omega_g = 0.004 \pm 0.002$ at $z = 3$, Since we know the physical state of all gas particles in the simulation, we can compute $\Omega_g$ directly by adding up the mass in cold, collapsed gas, i.e. gas with $\rho/\bar{\rho} > 1000$ and $T < 30,000$K. We find $\Omega_g \approx 0.0065$, 0.0036, and 0.0017 at $z = 2$, 3, and 4, respectively.

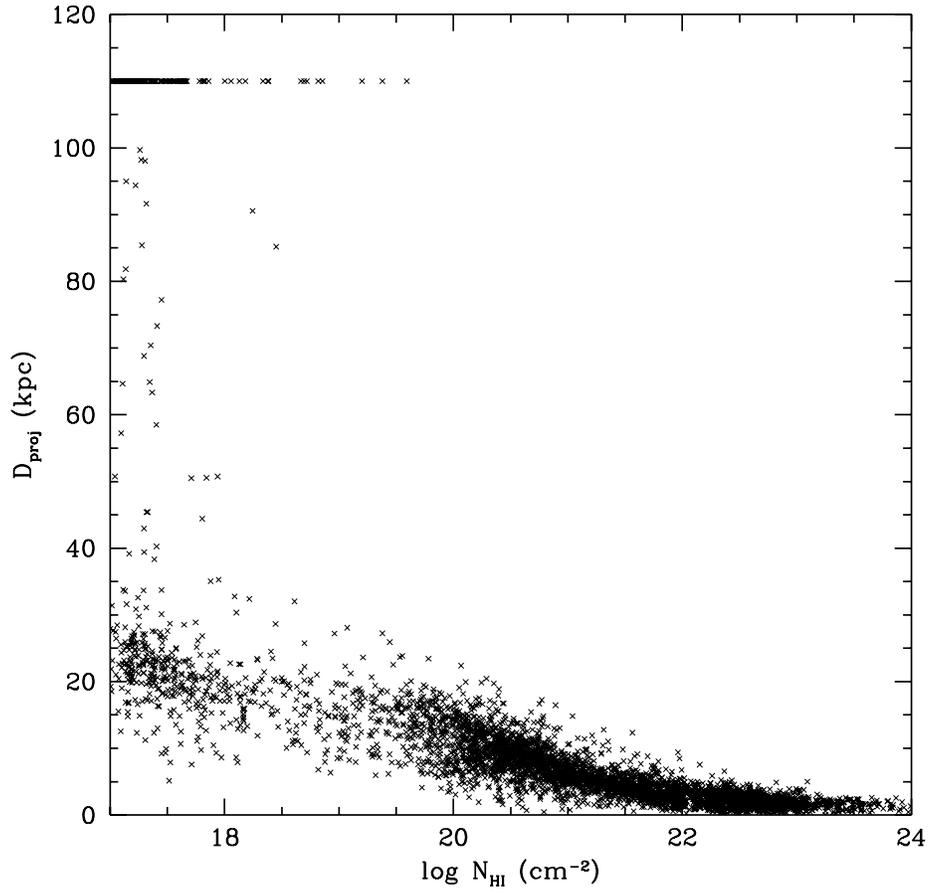

Fig. 2.— Projected distance $D_{\mathrm{proj}}$ between a line of sight with absorption column density $N_{\mathrm{HI}}$ and the nearest galaxy in the simulation. Separations greater than 100 kpc are plotted as 110 kpc.

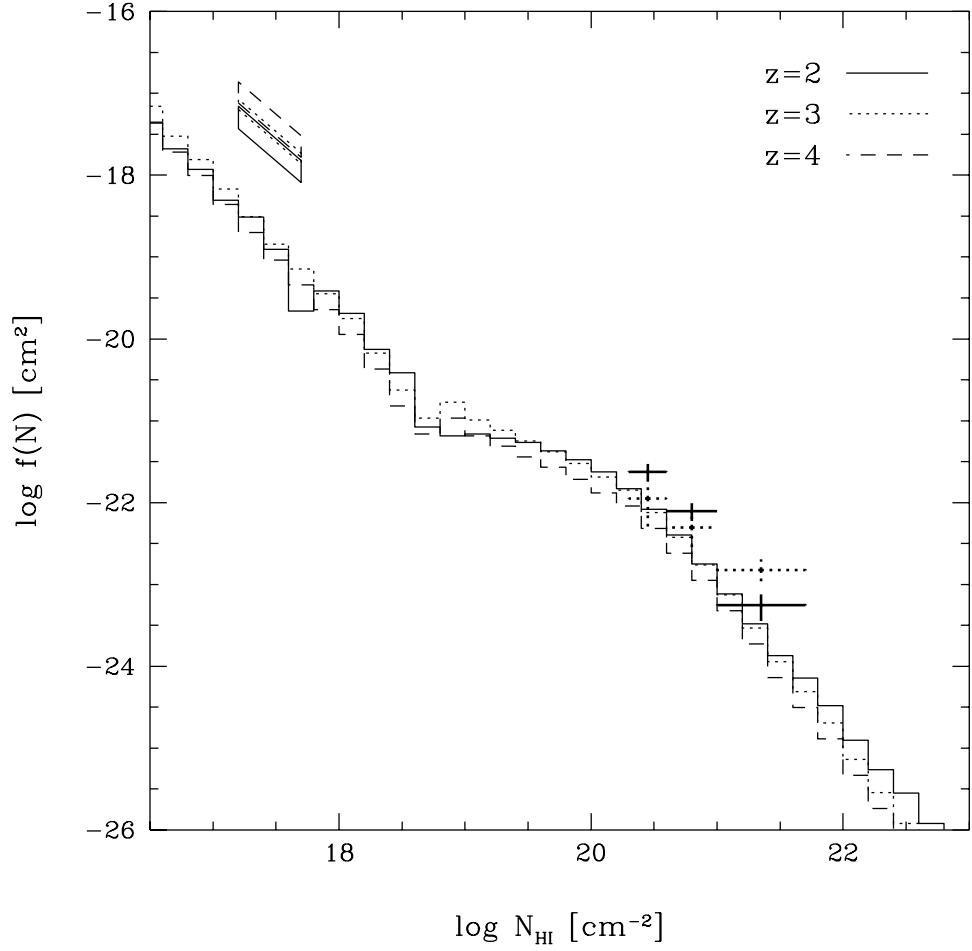

Fig. 3.— Logarithm of the frequency distribution $f(N, z)$ vs. logarithm of the neutral hydrogen column density. Histograms show the simulation results at $z = 2$ (solid), $z = 3$ (dotted), and $z = 4$ (dashed). Error crosses show observational results for damped sytems at $z = 2$ and $z = 3$, and diagonal boxes show observational constraints derived from the cumulative numbers of Lyman limit systems at $z = 2, 3$, and 4.

At first glance it is surprising that our $f(N, z)$ histograms lie below the observations but our values of $\Omega_g$ are similar to or even larger than the Wolfe et al. (1995) values. However, since the $f(N)$ distribution has a power-law index $\sim -1.7$, the contribution to $\Omega_g$ is dominated by the highest column density systems, and our directly measured $\Omega_g$ includes gas with $N_{\rm HI}$ above the $10^{21.7}\,{\rm cm}^{-2}$ upper bound used by Wolfe et al. (1995). The integration and extrapolation required to infer $\Omega_g$ from the observations makes it difficult to disentangle observational/statistical issues from physical issues (the cosmological model, the effects of star formation).

Wolfe (1988) and Schiano, Wolfe & Chang (1990) suggest that damped Ly$\alpha$ absorption arises predominantly in large ($\sim 50$ kpc) HI disks — the large size is needed to explain the observed incidence of absorption if the space density of damped systems is similar to that of present day disk galaxies. Our simulation reproduces the observed values of $f(N, z)$ and $\Omega_g$ with objects that are only $\sim 10$–20 kpc in radius because the number of systems able to produce damped Ly$\alpha$ absorption far exceeds the number of $L_\star$ galaxies at $z = 0$. Some of these systems will likely evolve into dwarf galaxies, while others will merge to form the smaller number of larger galaxies present today. We shall be able to address this issue more quantitatively once we have evolved the simulation to $z = 0$.

As shown in Figure 3, the number of Lyman limit absorbers falls short of the observed abundances by almost a factor of 10 at $N_{\rm HI} \approx 10^{17}\,{\rm cm}^{-2}$. Again, there is uncertainty caused by the small box size, but it seems likely that this discrepancy is statistically significant and could represent a failure of the CDM model we consider here. However, since many systems are marginally resolved, it may be that a modest increase in resolution would produce a large increase in the abundance of Lyman limit systems. We have compared this simulation to a run with the same initial conditions but 1/8 as many particles, and we find that the lower resolution has its most severe effect on the least massive radiatively cooled objects (Weinberg et al. 1995), which contribute much of the absorption near $N_{\rm HI} = 10^{17}\,{\rm cm}^{-2}$. Alternatively, we could be missing an additional population of absorbers produced by physical processes far below our resolution limit, such as fragmentation of cooling gas into pressure-confined clouds in galactic halos (e.g. Fall & Rees 1985; Wang 1993; Mo 1994).

A UV background field with $J_0 = 10^{-22}\,{\rm erg\,s^{-1}\,cm^{-2}\,sr^{-1}\,Hz^{-1}}$ is less intense than most estimates of the expected background from QSOs (e.g. Meiksin & Madau 1993). We would not expect a stronger background to have much effect on the abundance of damped systems, since these would still be self-shielded and mainly neutral. However, a stronger background would substantially reduce the abundance of Lyman limit systems, worsening the conflict with observations. Conversely, increasing $\Omega_b$ would increase the incidence of absorption by raising gas densities, neutral fractions, and cooling rates.

Despite the many limitations of our numerical modeling, we have been able to show that the CDM model considered here reproduces important features of the observations: the column density distribution (with the discrepancies already discussed) and the association between absorbers and galaxies. Equally important, we have demonstrated that it is now possible to predict the occurrence of hydrogen absorption directly from numerical simulations. This approach eliminates much of the uncertainty present in traditional semi-analytic calculations, so observations of damped Ly$\alpha$ and Lyman limit systems can provide more robust constraints on cosmological theories.

Although we have so far simulated only one cosmology, $\Omega = 1$ CDM with $\sigma_{16} = 0.7$, we can draw at least one more general conclusion. A model having much less power on the scales of our simulation would produce even fewer absorption systems and would be in greater conflict with observations, especially if the formation of molecular gas or stars has depleted the atomic hydrogen. It seems likely, therefore, that any theory able to explain the amount of neutral gas in observed damped Ly$\alpha$ systems must have more small scale power than $\sigma_{16} = 0.7$ CDM and/or a baryon density $\Omega_b > 0.05$. This requirement may eliminate some theories that can otherwise provide a good account of galaxy clustering and fluctuations in the microwave background.

We acknowledge helpful discussions with Craig Hogan, Martin Rees, Tom Quinn, and Art Wolfe. We extend special thanks to Ken Lanzetta for supplying us with his latest observational data. This work was supported in part by the Pittsburgh Supercomputing Center, the National Center for Supercomputing Applications (Illinois), the San Diego Supercomputing Center, the Alfred P. Sloan Foundation, NASA Theory Grants NAGW-2422, NAGW-2523, and NAG5-2882, NASA HPCC/ESS Grant NAG 5-2213, NASA Grant NAG-51618 and the NSF under Grant ASC 93-18185 and the Presidential Faculty Fellows Program. DHW acknowledges the support of a Keck fellowship at the Institute for Advanced Study during early phases of this work, and JM also acknowledges the W. M. Keck Foundation for support.